\documentclass[cha,twocolumn,amsmath, amssymb,graphicx,superscriptaddress]{revtex4-1}

\usepackage{bm}
\usepackage{ placeins}
\usepackage{subfigure}
\usepackage{graphicx}
\usepackage{amsthm}
\usepackage{notes2bib}
\usepackage{amsmath}
\usepackage{amssymb}
\usepackage{url}
\usepackage{enumerate}
\usepackage{epstopdf}
\usepackage{color}
\usepackage{mathtools}          
\usepackage{setspace}

\usepackage[usenames,dvipsnames,svgnames,table]{xcolor}
\usepackage{color}

\newcommand{\allblack}{\color{black}{}}

\newcommand{\R}{{\mathcal{R}}}

\newif\ifincludeXX
\includeXXtrue  

\begin{document}

\title{Low-dimensional paradigms for high-dimensional hetero-chaos}

\author{Yoshitaka Saiki}
\affiliation{Graduate School of Business Administration, Hitotsubashi University, Tokyo 186-8601, Japan}
\affiliation{JST, PRESTO, Saitama 332-0012, Japan}
\affiliation{University of Maryland, College Park, Maryland 20742, USA}
\author{Miguel A.F. Sanju\'{a}n}
\affiliation{University of Maryland, College Park, Maryland 20742, USA}
\affiliation{Nonlinear Dynamics, Chaos and Complex Systems Group, Departamento de F\'{i}sica\\ Universidad Rey Juan Carlos, Tulip\'{a}n s/n, 28933 M\'{o}stoles (Madrid) Spain}
\author{James A. Yorke}
\affiliation{University of Maryland, College Park, Maryland 20742, USA}

\allblack

\begin{abstract} The dynamics on a chaotic attractor can be quite heterogeneous, being much more unstable in some regions than others.
Some regions of a chaotic attractor can be expanding in more dimensions than other regions.
Imagine a situation where two such regions and each contains trajectories that stay in the region for all time – while typical trajectories wander throughout the attractor.
Such an attractor is ``hetero-chaotic'' (i.e. it has heterogeneous chaos) 
if furthermore arbitrarily close to each point of the attractor there are
points on periodic orbits that have different unstable dimensions. 
This is hard to picture but we believe that most physical systems possessing a high-dimensional attractor are of this type.
We have created simplified models with that behavior to give insight to real high-dimensional phenomena.
\end{abstract}

\date{\today} 

\maketitle
{\bf
Prediction and simulation
for chaotic systems occur throughout science. 
Predictability is more difficult when the ``chaotic attractor'' 
is heterogeneous, i.e. if different regions of the chaotic attractor are unstable in more directions than in others.
More precisely, when arbitrarily close to each point of the attractor there are different periodic points with different unstable dimensions, we say the chaos is {\bf heterogeneous} and we call it {\bf hetero-chaos}.
Simple illustrative models of hetero-chaos have been lacking in the literature, and here we present the simplest examples we have found.
\allblack
}

\section{Introduction} 

Predictability is especially difficult when a trajectory enters a region that has more unstable directions than the region it is leaving.
This appears to occur in geomagnetic storms and solar flares \cite{pariat_2017} or natural hazards \cite{guzzetti_2016} or earthquakes \cite{tian_2017} or weather \cite{patil_2001}. 
In such cases ``shadowing'' breaks down: numerical simulations no longer reflect true behavior.
In our work with simple whole earth weather models  (e.g., \cite{patil_2001}), the phase space had dimension $3$$\times$$10^6$, trajectories were chaotic, and we estimate that there were $3$$\times$$10^4$ unstable directions, that is, a tiny ellipse around an initial point would expand in $3$$\times$$10^4$ dimensions. 
The unstable dimension is usually about one-hundredth of the dimension of the dynamical system.
For storm conditions the regional unstable dimension is higher and thus prediction and simulation and data assimilation are much more difficult. 
\\
\indent
If the approximate state of the weather is known near some point $q$ in phase space,  then
after a short time, perhaps a few hours, the possible weather states lie on an expanding ellipse of some dimension $D$. We call $D$ the unstable dimension at $q$. 
To update the current state of the weather every few hours, it suffices to have enough observations to determine the location of the current state on that ellipsoid. The number of data observations -- point measurements of temperature, humidity, pressure, etc at nearby locations -- needed for that is proportional to $D$ which can be far smaller than the dimension of the state space.

For a barotropic atmospheric model
Gritsun \cite{gritsun_2008,gritsun_2013} found many unstable periodic orbits, and he found a wide variation in their numbers of unstable directions, all coexisting in the same system. He did not attempt to verify that these orbits were in the attractor. \\
\allblack
\ifincludeXX
\begin{figure}
\includegraphics[width=.45\textwidth]{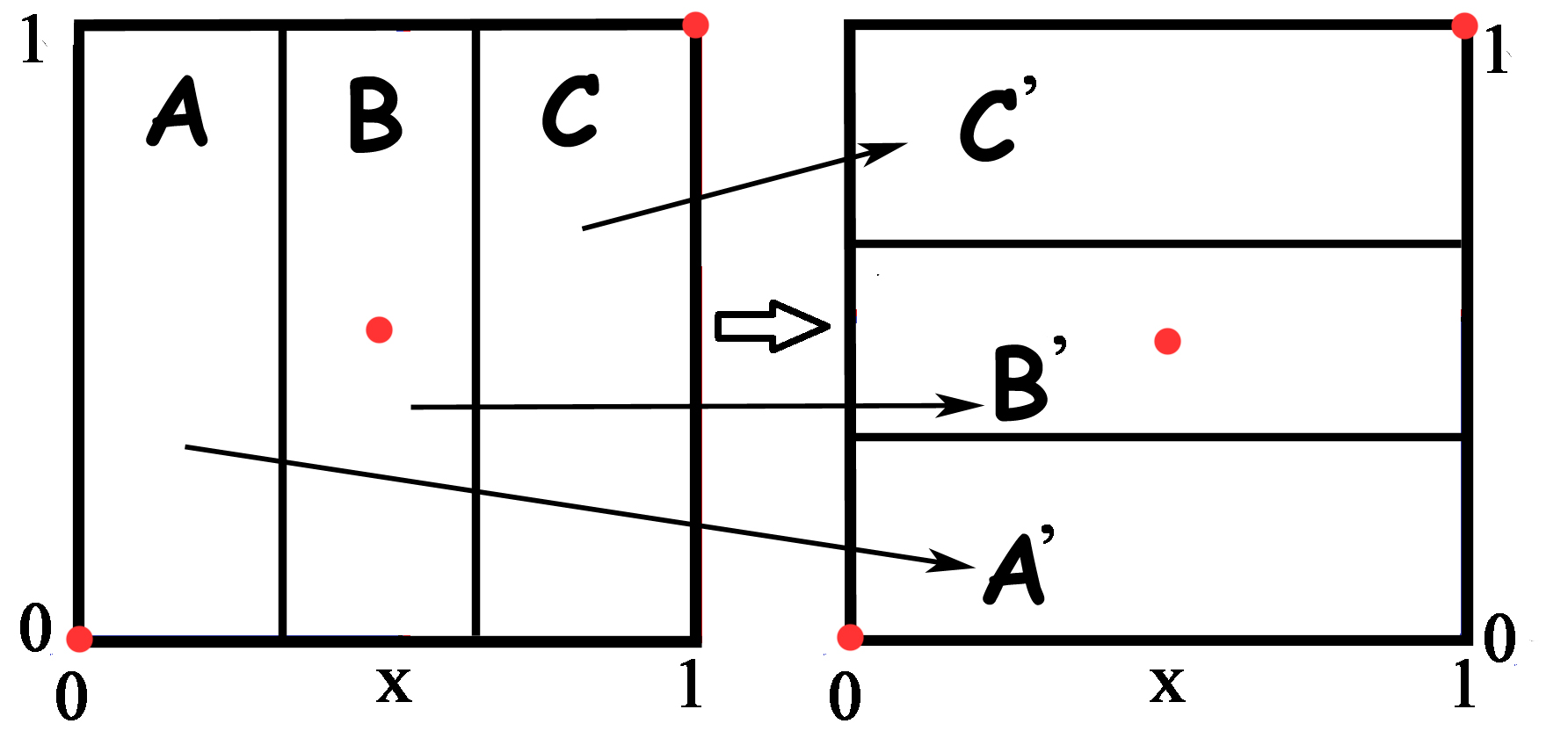}
\includegraphics[width=.45\textwidth]{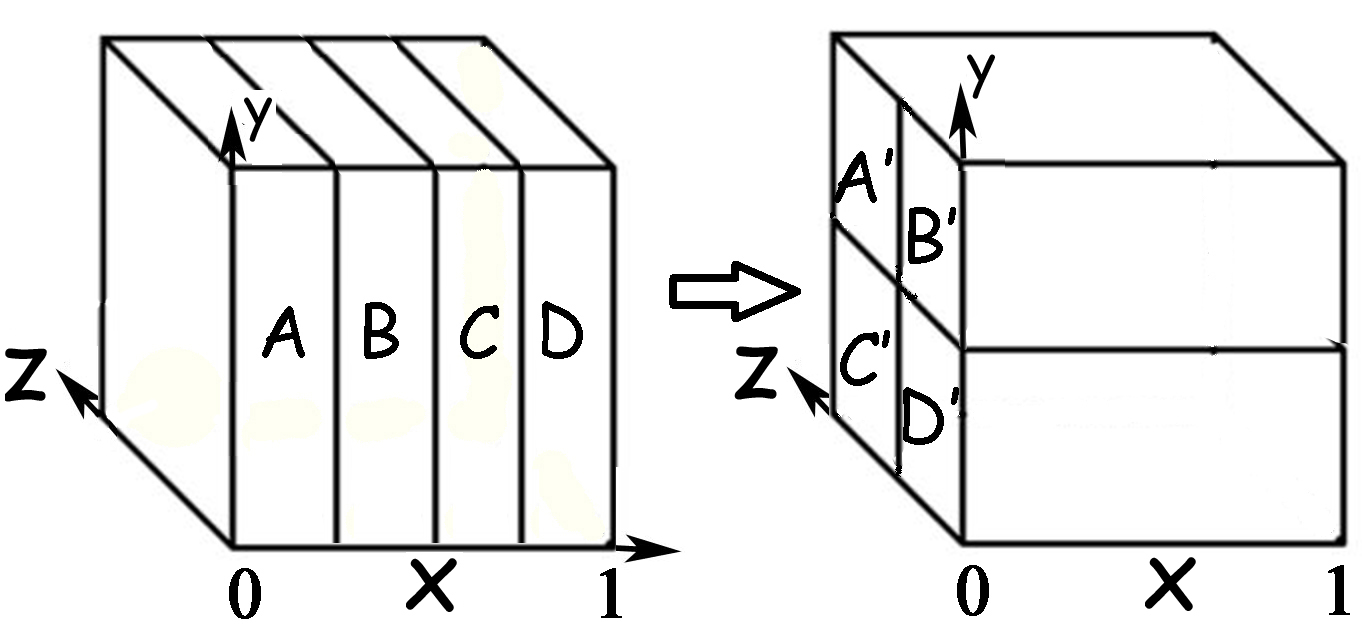}
\caption{ {\bf The homogeneous chaotic baker map in 2D and 3D.}
{\bf Top panel:}
The standard, i.e. 2D, baker map is defined by splitting 
the square into $p$ equal-sized vertical slabs where $p=3$ here. 
The square is mapped to the square, with
$x\mapsto 3x \bmod1$. 
Each vertical slab on the left maps to a different horizontal slab on the right, each stretched in the horizontal direction and shrunken vertically. The assignment of which maps to which has been chosen arbitrarily but the assignment is fixed. This map is homogeneously chaotic with one expanding direction.  Each slab has a fixed point denoted by a red dot. (Note that two are on slab boundaries). The images are denoted by primes~{\bf $\prime$} so for example
$A$ maps onto $A'$. 
{\bf Bottom panel:}
We provide a 3D version by slicing a unit cube into four equal-sized pizza-box shaped slabs as shown.
The cube is mapped to the cube, using
$x\mapsto 4x \bmod1$ so each pizza-box is expanded to the width of the cube; and
each pizza-box maps to a different shoe box-shaped region on the right, each shrunken by a factor of two in the $y$ and $z$ coordinates. 
The assignment of which pizza-box maps to which shoe box has been chosen arbitrarily but the assignment is fixed. This map is homogeneously chaotic with one expanding direction.
\allblack }
\label{fig:baker}
\end{figure}
\fi
\ifincludeXX
\begin{figure}
  \includegraphics[width=.45\textwidth]
{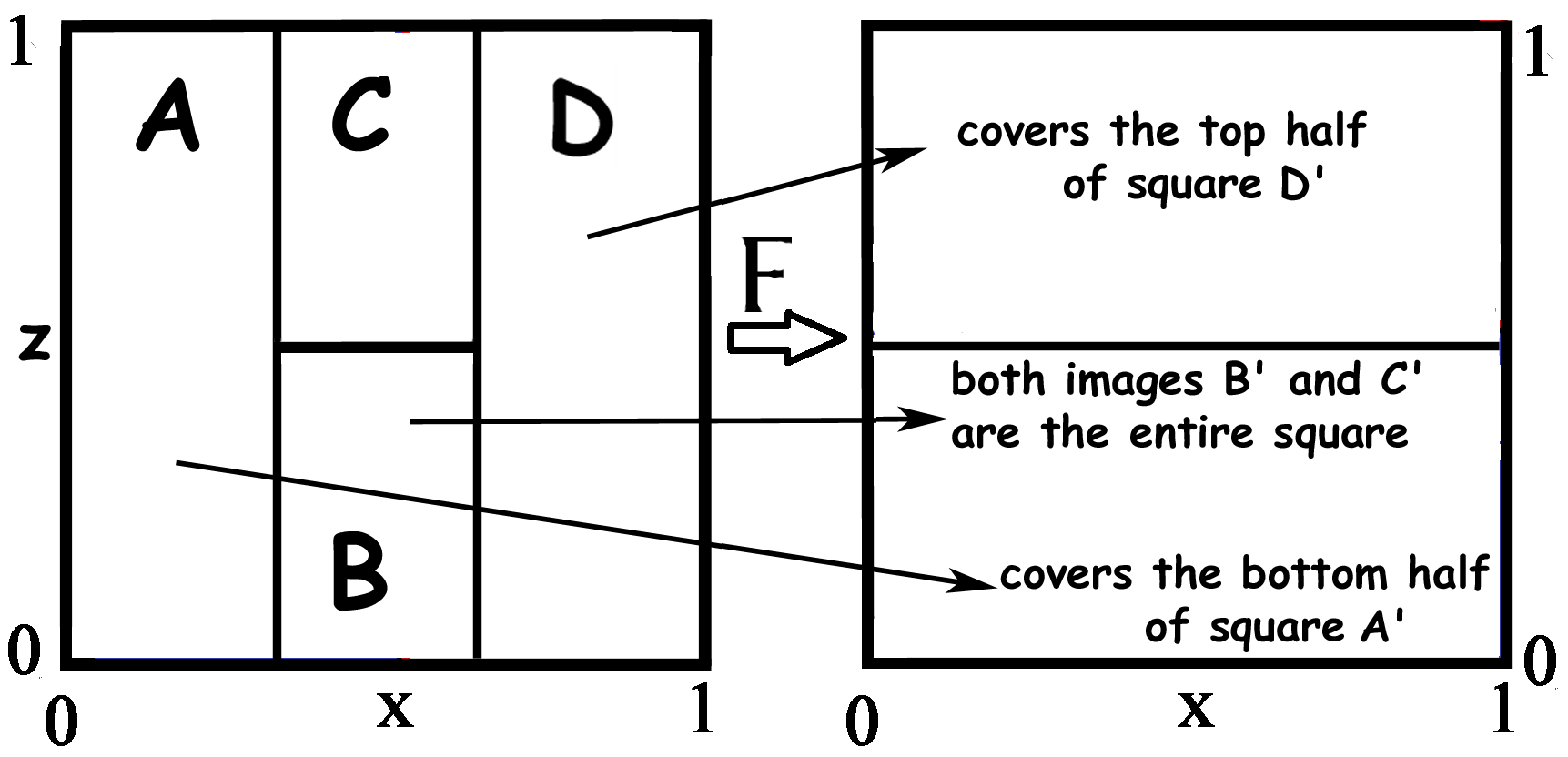}
  \caption{
    {\bf Our 2D Hetero-chaotic (HC) baker map $B_{HC}(x,z)$}. 
The figure shows a three-piece version of the baker map. 
We divide $0\le x<1$ into three intervals,
 $L=[0,1/3),M=[1/3,2/3),$ and $R=[2/3,1)$,  and divide the square into $3$ tall rectangles 
 $A, B\cup C,$ and $D$
 whose bases are $L, M,$ and $R$. 
The map  $B_{HC}$ is defined as follows:
For $x\in L,~z\mapsto z/2.$ 
For $x\in M,~z\mapsto 2z \bmod 1.$ 
For $x \in R,~z\mapsto z/2 + 1/2.$ 
And then $x\mapsto 3x \bmod1$. 
Hence $B_{HC}$ expands each rectangle horizontally to full width as shown. The region $\R_1 = A\cup D$ is contracted vertically. The region 
$\R_2=B\cup C$ is expanded in both coordinates  so that the images of $B$ and $C$ each cover of the entire square. 
Hence $\R_1$ and $\R_2$ are regions of one- and two-dimensional instability. 
}
\label{fig:2D_HC_baker}
\end{figure}
\fi
\ifincludeXX
\begin{figure}
  \includegraphics[width=.45\textwidth]
{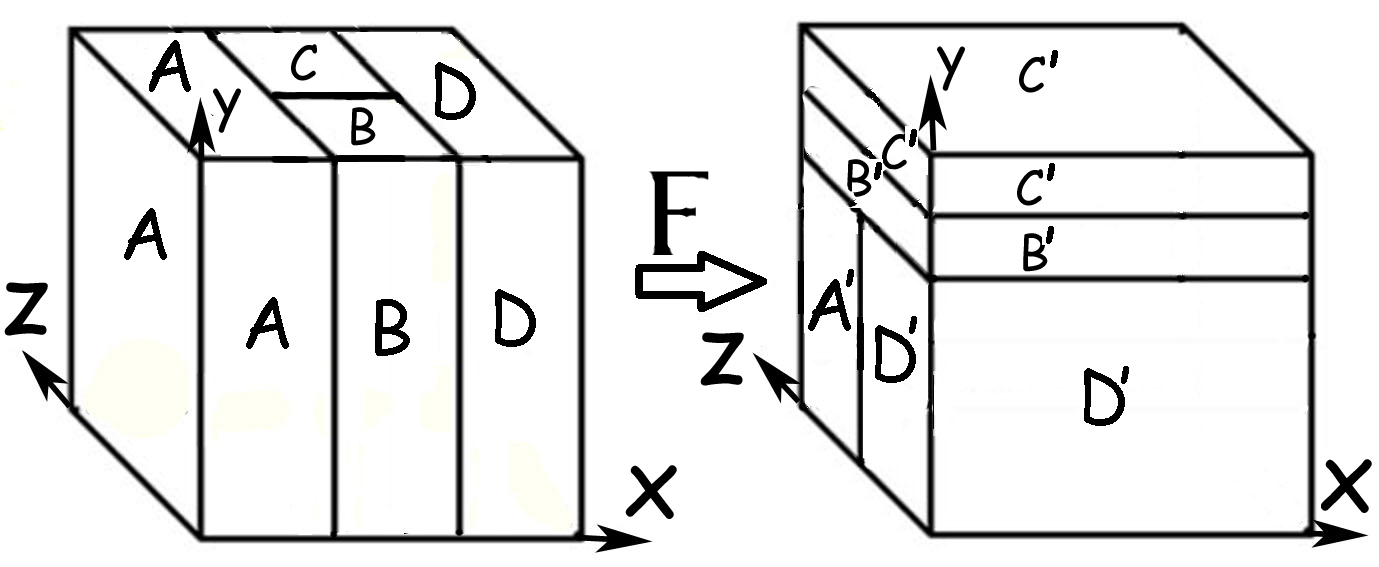}
  \caption{
    {\bf A volume-preserving 3D version of Fig.~\ref{fig:2D_HC_baker}}. 
    Here the $X$-$Z$ plane plays the role of $X$-$Z$ in Fig.~\ref{fig:2D_HC_baker} and the $Y$ coordinate has been added.
Here the cube is partitioned into four regions  $A$, $B$, $C$, and $D$ 
and for all four
 $x~\mapsto~3x~\bmod1$,
and each is mapped into a region of the same volume. 
We write $X'$ for the image of any region $X$.
Both $B$ and $C$ expand in two directions and contracting in one, both having $1/6$ the volume of the cube. $A$ and $D$ each have volume $1/3$ and expand in only the $x$ direction. 
In other words, the $y$-height of $A'$ and $D'$ is $2/3$ and the $y$-height (or thickness) of $B'$ and $C'$ is $1/6$.
Note that under $F$ the $y$ coordinate shrinks for all four regions.
 }\label{fig:3D_HC_baker}
\end{figure}
\fi 
\ifincludeXX
\begin{figure}
  \includegraphics[width=.23\textwidth]
{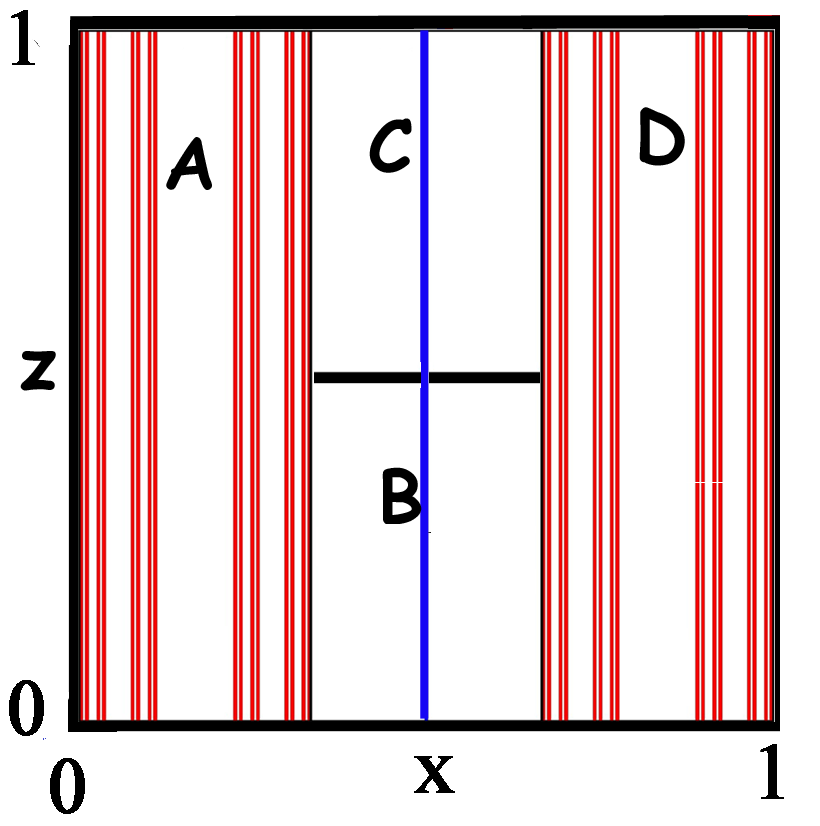}
  \caption{
    {\bf The invariant ``index'' sets}.
For the map in Fig.~\ref{fig:2D_HC_baker}, the vertical red lines here constitute the invariant set whose trajectories stay in  $\R_1$ and the vertical blue line (at $x=1/2$) is the invariant set of points whose trajectories stay inside  $\R_2$. }
\label{fig:Baker-index-sets}
\end{figure}
\fi
\indent
{\bf  Baker map.}
Our first examples with hetero-chaos are based in part on 
the well-known ``baker map''. It was defined in 1933 by Seidel \cite{seidel_1933}. The map is defined by dividing the square into $p$ equal vertical strips. Seidel used $p=10$. We use $p=3$ in Fig.~\ref{fig:baker} and $p=2$ is most common in the literature.
Each strip is mapped to a horizontal strip by squeezing it vertically by the factor $p$ and stretching it horizontally by the same factor. The resulting horizontal strips are laid out covering the square. 

We also show a three dimensional version. In both of these baker maps, the unstable dimension $D$ is $1$ and in particular is constant. In such cases we call the chaos {\bf homogeneous}, and refer to it as {\bf homogeneous chaos.}
The baker maps are area or volume preserving. 
The earliest use of the map name ``baker'' that we have found appears in the 1956 {\it Lectures
on Ergodic Theory} by Paul Halmos \cite{halmos_1956}. 
He writes that the actions of the map are reminiscent of the kneading dough and writes that it is 
``sometimes called the baker's transformation''. 

In the bottom half of Fig.~\ref{fig:baker} we give a 3D baker map. Here the unstable dimension is $D=1$ (and the map contracts the $y$ and $z$ directions). To convert this example into one with unstable dimension $D=2$ and stable dimension 1, just take the inverse, mapping each box $X'$ on the right to the box $X$ on the left. 
For area-contracting (``skinny'') and area-expanding (``fat'') 2D-baker maps, see \cite{farmer_1983} and \cite{alexander_1984}, respectively.

{\bf Our hetero-chaotic baker maps.}
The baker maps in Fig.~\ref{fig:baker} are homogeneously chaotic, but we here modify them to be hetero-chaotic (HC). 
We introduce two such modified maps
in Figs.~\ref{fig:2D_HC_baker} and ~\ref{fig:3D_HC_baker}
as prototypes for understanding attractors with far higher dimension. 

\allblack
In Fig.~\ref{fig:2D_HC_baker},
there are two regions ($\R_1 = A\cup D$, the left and right thirds of the square) where the dynamics is unstable in one direction (the $x$ coordinate) while in the middle third ($B\cup C$), denoted $\R_2$, 
it is unstable in both $x$ and $z$ coordinates. 
See Fig.~\ref{fig:3D_HC_baker} for a 3D invertible 
volume-preserving version.
There exist trajectories that stay in either region but almost every trajectory wanders through the entire square.
See Fig.~\ref{fig:Baker-index-sets} for the homogeneously chaotic invariant set formed by such limited trajectory. We call the set an index set, as described later.  
For simplicity, we ignore the dynamics of all points on the boundaries of the rectangles $A,B,C,D$. 
In the example, in $\R_1$ the map contracts the $y$ direction by a factor of $2$ while it expands by a factor of $2$ in $\R_2$. 
Hence a periodic orbit that has most of its points in $\R_1$ will have unstable dimension $1$ while if most are in $\R_2$ it has unstable dimension $2$.

\allblack
 {\bf Unstable Dimension Variability.}
If a periodic orbit is unstable in $k$ directions, 
we say {\bf it has UD-$k$}. 
In our 2D examples, UD-$1$ orbits are saddles and UD-$2$ orbits are repellers. Hence if an attractor (with a dense trajectory) has a UD-$1$ orbit and a UD-$2$ orbit, the attractor has UDV.

When an attractor has 2 periodic orbits that are unstable in different numbers of dimensions, we say the attractor has {\bf Unstable Dimension Variability (UDV)} \cite{kostelich_1997}.

\bigskip
{\bf Conjecture~1.} Almost every chaotic attractor has the property that if there is one UD-$k$ periodic orbit, then 
there are infinitely many UD-$k$ periodic points and they lie arbitrarily close to each point of the attractor. 

\section{Hetero-chaos}
A set $S$ is a {\bf chaotic attractor} if (1) it is {\bf invariant} (i.e., if a trajectory is in $S$ at some time, then it is in $S$ for all later time), (2) $S$ has a dense trajectory with at least one positive Lyapunov exponent, and (3) trajectories near $S$ are attracted to it as time increases.

We will say a chaotic attractor has {\bf hetero-chaos} if arbitrarily close to each point of the attractor there are  periodic points on UD-$k$ periodic orbits and this is true for multiple values of $k$.
In  \cite{das_2017a} it is called ``multi-chaos'' but ``hetero-chaos'' seems more appropriate.
We expect that most high-dimensional attractors are hetero-chaotic.

   A consequence of UDV is that any trajectory that wanders densely through the invariant set will occasionally get very close to each periodic point. Therefore that trajectory will spend arbitrarily long intervals of time near each of the fixed points (or periodic orbits).  
Hence for each time $T>0$ the trajectory's time-$T$ positive Lyapunov exponents
will occasionally be  the same as for the periodic orbit it approaches. 

\bigskip
{\bf Conjecture~2.} UDV always implies hetero-chaos.

\bigskip

{\bf Results for the hetero-chaos baker maps in Figs.~\ref{fig:2D_HC_baker} and ~\ref{fig:3D_HC_baker}.}
We can prove the maps in Figs.~\ref{fig:2D_HC_baker} and ~\ref{fig:3D_HC_baker} are hetero-chaotic. 
Specifically, arbitrarily close to each point $q$ in the square there are periodic points of different UD-$k$.

{\bf Degenerate periodic orbits.} It is possible for some periodic orbits to be degenerate. For our 2D HC-baker map, a simple period-2 example has $x=1/8$ and $3/8$. 
Then for each $z\in[0,1)$, the point
$(1/8,z)$ maps to $(3/8, z/2)$ which maps to $(1/8,z)$, so this is periodic. Clearly there is an infinite collection of such period-2 orbits. 
There is a corresponding family in the 3D version of the map. 
More generally, at each iterate of a trajectory, nearby points differing only in the $z$-coordinate either move apart by a factor of 2 or move closer by a factor of 2, and if a periodic orbit has an equal number of both types, then the orbit is neutrally stable in the $z$ direction. 
All such degenerate orbits have even period. Non-degenerate orbits are called {\bf hyperbolic}.

{\bf Counting hyperbolic periodic orbits.}
The numbers of period-$N$ hyperbolic UD-1 and UD-2 periodic orbits are both approximately $3^N$ when $N$ is large.

{\bf Ergodicity.}
Our 2D and 3D HC-baker maps (denoted by $F$ below) are ``ergodic'' in the following sense. For every continuous function $\phi$ on the square or the cube, write $\hat\phi$ for the average value of $\phi$ on the cube or the square. The map $F$ is {\bf ergodic} if 
for almost every initial point $q$
 the trajectory average 
$$
\frac{1}{N}\sum_{n=1}^N \phi(F^n(q)) \to \hat\phi.
$$

Due to the ergodicity we can also conclude that there is a dense trajectory. In fact ergodicity for our baker maps implies that for almost every initial point $q$, the trajectory 
$F^n(q)$ for $n\ge0$ comes arbitrarily close to every point of the square or cube, respectively.

The proofs of the statements that $F$ is hetero-chaotic and ergodic will be provided elsewhere.

{\bf The route to hetero-chaos when the attractor changes continuously with a parameter.}
 In addition to presenting low-dimensional examples, the purpose of this paper is ask how hetero-chaos arises from homogeneous chaos as some physical parameter is varied. 
 We show numerical evidence that the zig-zag example in Fig.~\ref {fig:zigzag} (where $\sigma = 5$) 
 is homogeneously chaotic for $\alpha < \alpha_{HC} \sim 0.31$ and is hetero-chaotic for $\alpha > \alpha_{HC}$. Similarly we show numerical evidence that the Kostelich map (Fig.~\ref{fig:two_cantor_sets}) is homogeneously chaotic for $\sigma<\sigma_{HC}=1/\pi\sim 0.318$ 
 and is hetero-chaotic for $\sigma>\sigma_{HC}$, 
 when $\alpha=0.07$. 
\allblack

 We believe if an attractor is changing continuously, the transition will occur at a periodic orbit bifurcation and we give some examples of this transition.

{\bf The crisis route to  hetero-chaos.}
As some parameter, say $\alpha$, is varied, a ``crisis'' occurs  at some value $\alpha_0$ when there is a sudden discontinuous change in the size of a chaotic attractor. 
Hence, a crisis can be seen as a sudden jump in the plot of an attractor versus $\alpha$.
On the side of $\alpha_0$ where the attractor is small, the attractor could be homogeneously chaotic. On the other side, the attractor can be much larger and can include periodic orbits of a different UD value. Then the attractor is hetero-chaotic. See \cite{alligood_2006, das_2017a, viana_2005, pereira_2007}.

{\bf  The continuous route to  hetero-chaos.} 
If as a parameter $\alpha$ is varied,
a homogeneous chaotic attractor suddenly becomes hetero-chaotic after some $\alpha =\alpha_{HC}$, we say a   {\bf hetero-chaos bifurcation (HCB)} occurs at $\alpha_{HC}$. 
What is the nature of this bifurcation?
As a parameter changes, a periodic orbit in a chaotic attractor can migrate to a region that is more unstable, and the orbit's UD value can increase. Then an exponent of that orbit will pass through $0$ and a bifurcation will occur. 
Or a new pair of orbits can appear in an analogue of a saddle-repeller bifurcation, with UD values $k$ and $k+1$ for some $k>0$.

\bigskip
{\bf Conjecture~3.}  
 For a typical attractor, if an HCB 
occurs as the attractor changes continuously (without a 
crisis), then there 
will be 
a periodic orbit bifurcation, i.e.,
either period-doubling or pitchfork or Hopf or  pair-creation such as saddle-repeller. 
\bigskip

{\bf Expanding regions $\R_k$ and ``index sets''}
Let $\R_k$ denote the region of phase space in which the dynamics (specifically, the map's Jacobian) is $k$-dimensionally expanding; see e.g. Fig.~\ref{fig:2D_HC_baker}.
We call the largest invariant set that lies wholly in $\R_k$ the  {\bf index-$k$ set}.
In Fig.~\ref{fig:2D_HC_baker}, $\R_1$ and $\R_2$ are described. 
The index sets for the 2D HC baker map are shown in Fig.~\ref{fig:Baker-index-sets}.

At the center of 
Fig.~\ref{fig:two_cantor_sets}-Right, there is a different $\R_2$, the white rectangle ($1/3<x<2/3, -c <y<1+c$) where $c=(1-\alpha)/(\sigma-\alpha)\approx 0.13$, and $\R_1$ is the rest, excluding boundary points. 

 It probably seems strange that the existence of two periodic orbits with different UD values has such a dramatic consequence for an attractor that it implies hetero-chaos.
Our response is that these orbits generally lie in index sets, that can be quite big as Figs. 
\ref{fig:2D_HC_baker} and 
\ref{fig:two_cantor_sets} illustrate.

\section{Hetero-chaos connects many phenomena like fluctuating exponents (FE) and  UDV} 
Hetero-chaotic attractors contain periodic orbits with different UD values. A typical trajectory will  return near each, occasionally spending long times near them before moving on, and while near the periodic orbit of a region, it will have the same number of positive finite-time Lyapunov exponents (FTLEs) as the periodic orbit. As it moves among the periodic orbits, 
its number of positive FTLEs fluctuates (for each time $T>0$);  see \cite{dawson_1994, dawson_1996}. This property is referred to as FE (Fluctuating Exponents). 
Some papers have used the term UDV to mean FE.
UDV and FE are both implied by other dynamical phenomena in the literature such as riddled basins,
blowout bifurcations, on-off intermittency, and chaotic itinerancy \cite{ott_1993, platt_1993, heagy_1994, ott_1994, tsuda_2009}.

Transitions from homogeneous chaos to FE or UDV have been observed in \cite{dawson_1996, moresco_1997, barreto_2000b}, but the mechanism of the transitions is not discussed.

{\bf Shadowing.} It is important for a physicist to know how good a numerical simulation is -- as in a climate simulation -- and for how long it is valid. 
When each numerical trajectory stays close to some actual trajectory of the system, we say the system has the {\bf shadowing property,} i.e. simulations are realistic. 

When a trajectory moves from a region where the dynamics has fewer unstable directions to a region where it has more, shadowing fails, and trajectories become unrealistic
-- see Fig.~3 of 
 \cite{grebogi_2002}. Such a transition causes fluctuations in the number of positive FTLEs,
 which means FE will be common in  higher-dimensional attractors.

The FE property implies shadowing fails, as was established by   \textcite{dawson_1994}. 
Homogeneous chaotic systems can have the shadowing property but hetero-chaotic systems cannot, as shown for UDV in  \cite{sauer_1997,
yuan_2000, grebogi_2002}.

{\bf Hetero-chaos is not Hyper-chaos.}
Hetero-chaos should not be confused with ``hyper-chaos'' \cite{harrison_2000}.
A hetero-chaotic attractor can have one or more positive Lyapunov exponents. It need not be hyper-chaotic  (i.e., having more than one positive Lyapunov exponent). 
Furthermore all periodic orbits of a hyper-chaotic attractor might  have the same UD value, in which case it would not be hetero-chaotic.

{\bf UDV in the mathematics literature.}
The first examples of a (robust) invariant set 
containing periodic orbits with different UD values were given by \textcite{abraham_1970, simon_1972} in four and three dimensions, respectively.  ``Robust'' means the property persists under all sufficiently small perturbations. 
 Later it was mathematically studied using the notions of ``blenders'' and ``hetero-dimensional cycles'' (see \cite{bonatti_2005} and references therein).
That literature generally shows no interest in whether their invariant sets are (physically observable) attractors.

\ifincludeXX
\begin{figure}[t]%
\includegraphics[width=.23\textwidth]
{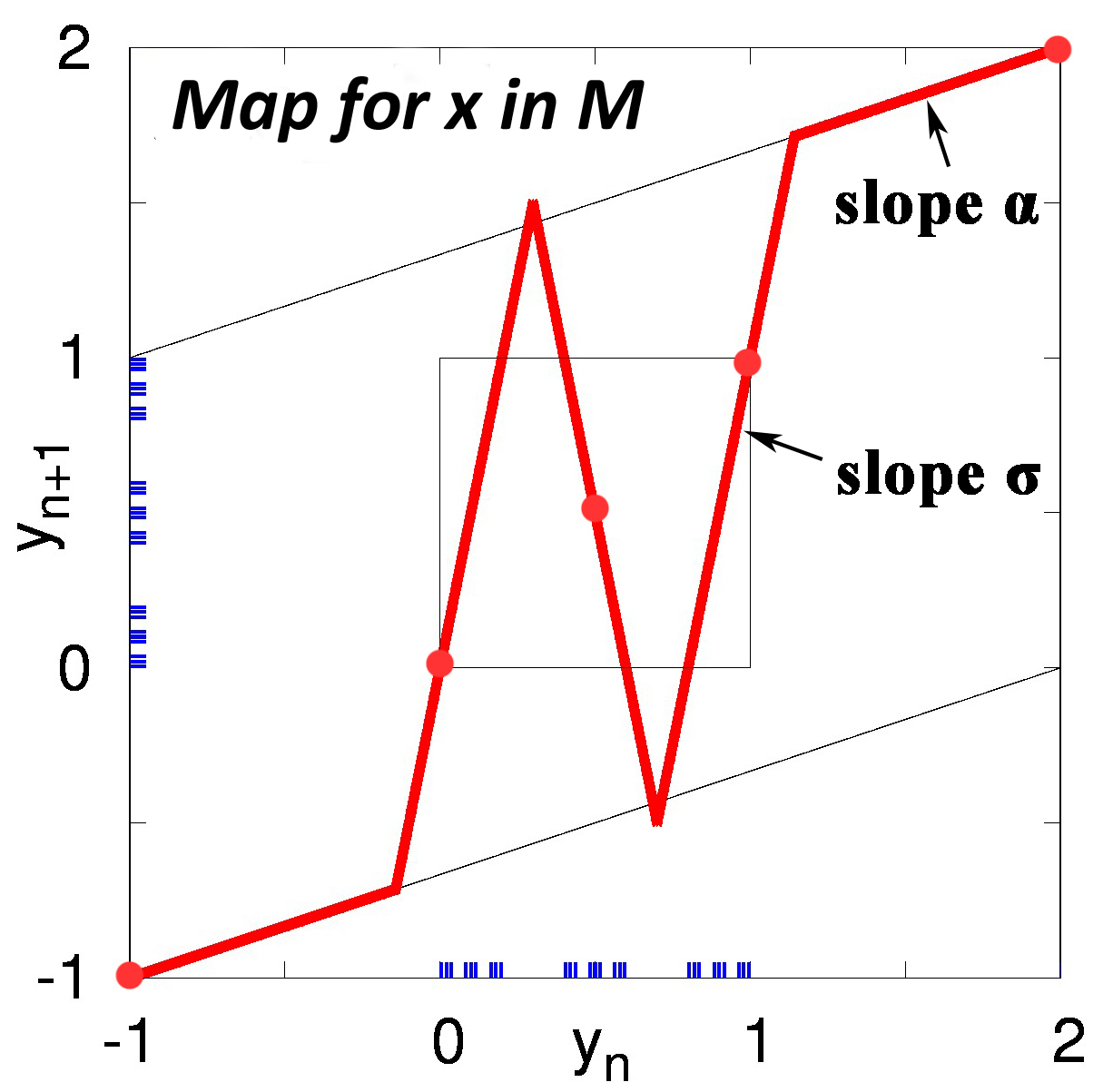}
\caption{{\bf Defining our Zigzag Map}. 
Here, as in Fig.~\ref{fig:2D_HC_baker}, 
the definition of the map depends on which of the three intervals $x$ is in: $L=[0,1/3),M=[1/3,2/3),$ and $R=[2/3,1)$. 
For $x\in L$,  
$y\mapsto -1+\alpha (y+1)$. 
For $x\in R$, 
$y\mapsto 2+\alpha (y-2)$. 
For $x\in M$, the 
figure shows the map. 
Each of the three maps is from [-1,2] into itself. 
The horizontal coordinate $x\mapsto 3x \bmod1$.
Each slope in the map shown is either $0<\alpha<1$ or $\pm\sigma$, where $\sigma>3$.
Here $\alpha = 1/3,$ and $\sigma=5$ both here and in Fig.~\ref{fig:two_cantor_sets}.
All 5 fixed points are shown with large red dots.
The Zigzag Map has an invariant fractal set on the vertical line for which $x=1/2$ ($x$ is not shown). Here these $y$ values are   illustrated using dots on axes.
}
\label{fig:zigzag}
\end{figure}
\begin{figure}	
  \includegraphics[width=.238\textwidth,height=.238\textwidth]
{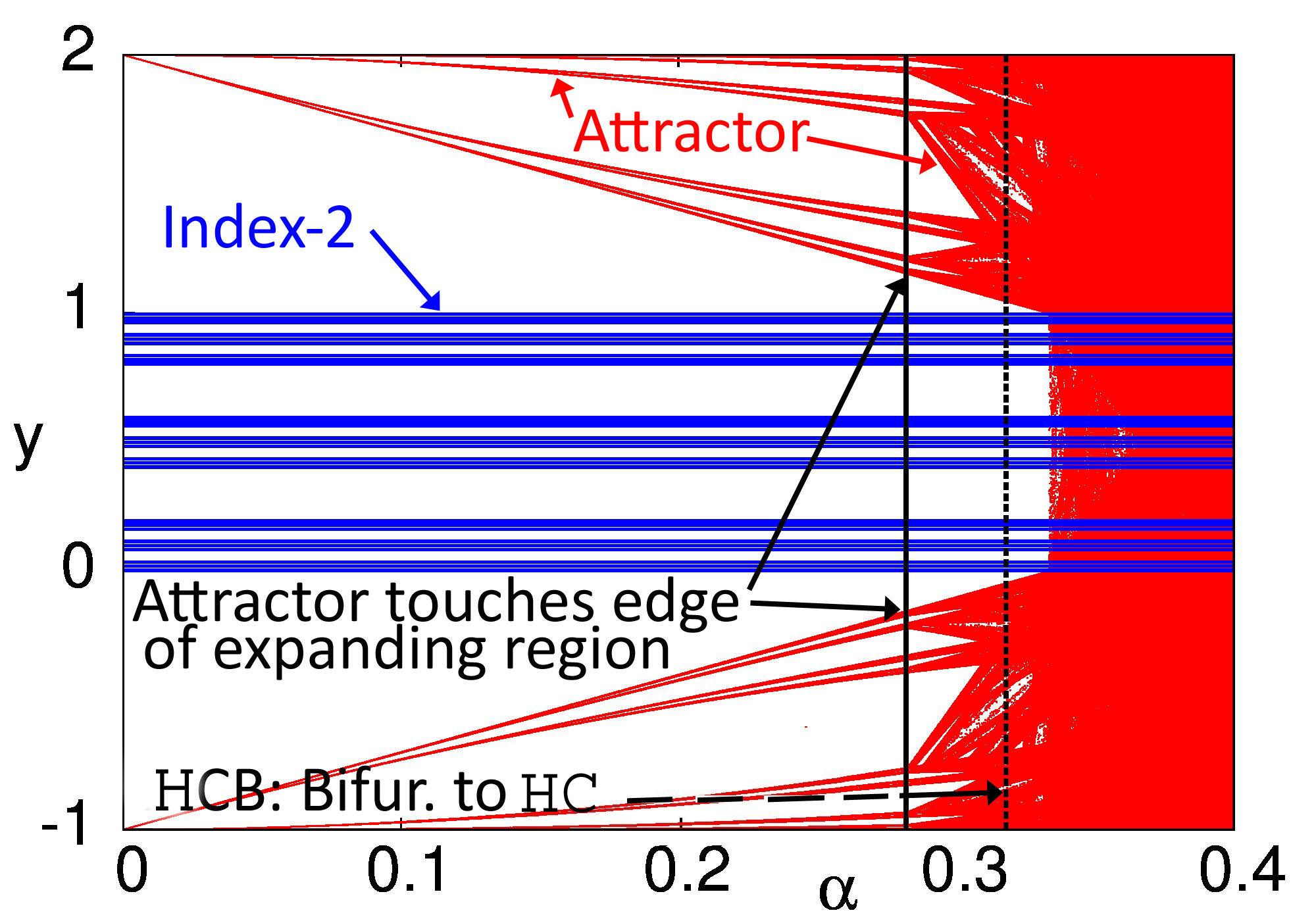}
  \includegraphics[width=.238\textwidth,height=.238\textwidth]
{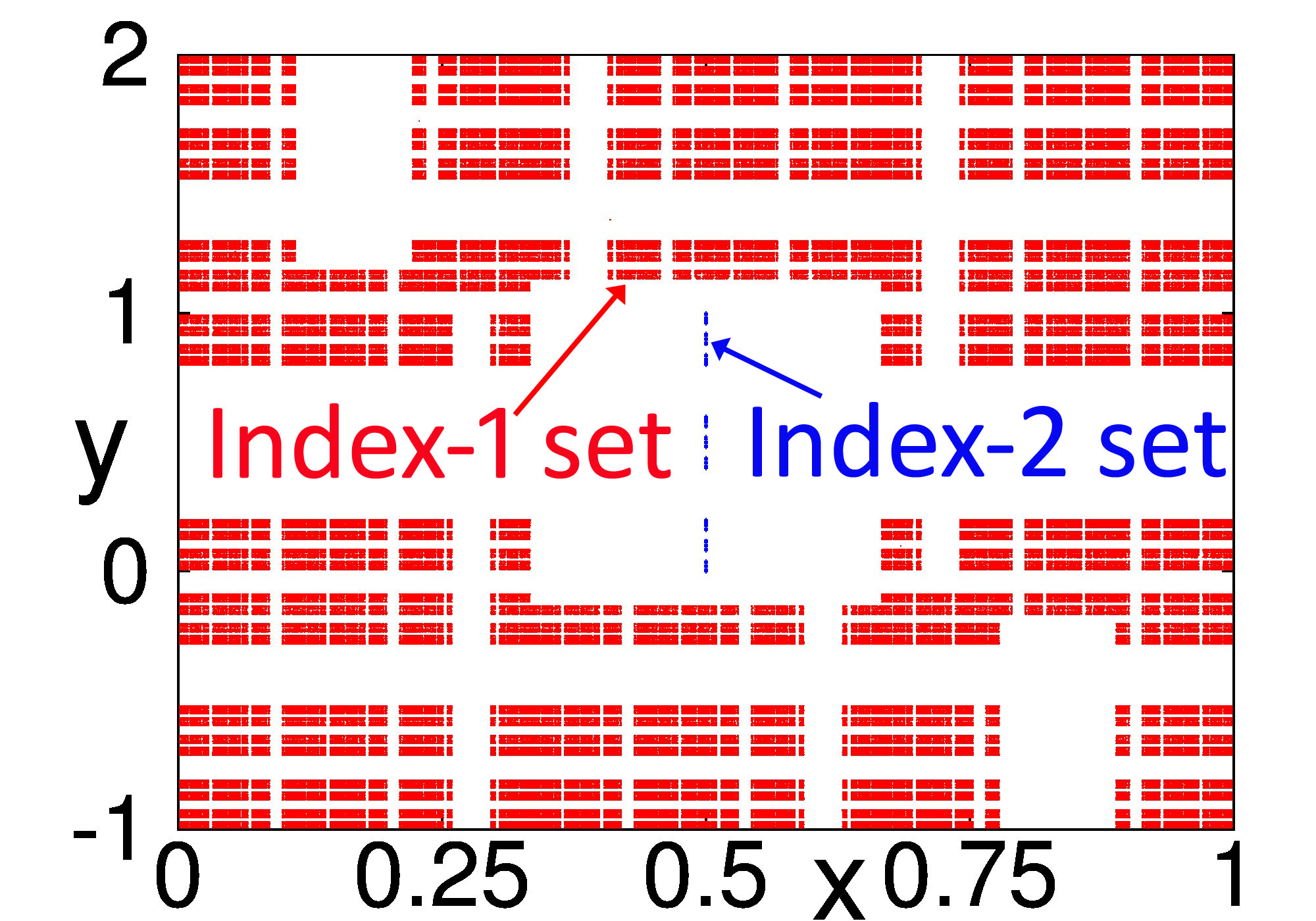}
  \caption{\textbf{The Zigzag Map's bifurcation diagram and index sets.}
{\bf Left panel.}  The chaotic attractor (red) is shown increasing in size as $\alpha$ increases.
The blue set is the index-$2$ set.     
At $\alpha\approx 0.28$ (solid black vertical line) the attractor begins to move into the expanding region, but the attractor does not contain repelling periodic points until after $\alpha_{HC}\approx 0.31$ (dotted black vertical line), when a period-4 repeller exists. Then there is hetero-chaos. 
    At $\alpha=1/3$ there is a ``crisis'' after which the attractor jumps in size and is the entire $x$-$y$ square.
{\bf Right panel.}   Here $\alpha = 0.4~ (> 1/3)$. We show only 
the index-$1$ set (red) and the index-$2$ set (blue), which is on the vertical line $x=1/2$. This illustrates that within the hetero-chaotic attractor (the entire square) there are relatively large index sets.
}\label{fig:two_cantor_sets}
\end{figure}
\fi


\section{Two more Hetero-Chaotic Maps}
{\bf Our ``Zigzag'' Map and its route to hetero-chaos.} 
As with the 2D HC Baker Map,  
the next 2D map has $x$ dynamics described by $x\to 3x ~\bmod 1$,
and its $y$ dynamics depends on whether $x$ is in $L$, $M$, or $R$. It has two slope parameters, $\alpha$ and $\sigma$. 
 Figure~\ref{fig:zigzag} shows the $y$ dynamics on $M$ and the caption gives the map also on $L$ and $R$.
The map has an index-$2$ fractal invariant set on the vertical line at $x=1/2$ for every $\alpha$ and every $\sigma>1$; (we use $\sigma = 5$ and then its
dimension is ${\ln 3}/{\ln 5}\approx 0.683$).
The attractor is chaotic for all $\alpha>0$, and for $\alpha<0.28$  is an index-$1$ set.

As $\alpha$ increases from $0$, at $\alpha_{HC}\approx 0.31$ (see
the left panel of Fig.~\ref{fig:two_cantor_sets}), there is   a pitchfork bifurcation of a period-$4$ periodic orbit, one of whose branches consists of repellers. Numerically this appears to be the first occurrence in the attractor of a repelling periodic orbit.
This observation supports Conjecture~3.
Hence the HCB occurs at $\alpha_{HC}$.

At $\alpha=1/3$, the attractor collides with the index-$2$ set, after which the attractor suddenly jumps in size, covering the whole $x$-$y$ square. 
For $\alpha=0.4$, the attractor is the whole torus and both index-$1$ and index-$2$ sets coexist 
(see right panel of Fig.~\ref{fig:two_cantor_sets}).
We have identified the index sets by using the Stagger-and-Step method \cite{sweet_2001c}.

{\bf Kostelich map.} The following smooth map  \cite{kostelich_1997,das_2017a} is defined on a two-dimensional torus:
\begin{equation*}
\begin{array}{lcl}
x_{n+1}&=&3 x_n \bmod1\\
y_{n+1}&=&y_n -\sigma \sin(2 \pi y_n)+\alpha (1-\cos(2 \pi x_n))\bmod1. 
\end{array}
\end{equation*}
It has an HCB whose periodic orbit bifurcation is a period-doubling at the origin,  a fixed point that becomes a repeller. We find numerically that immediately after the bifurcation  the chaotic attractor has a dense set of repellers and a dense set of saddles. This observation also supports Conjecture~3.
For $\alpha=0.07$ and $\sigma\in (0.2, 1/\pi) $, there is a chaotic attractor for which all periodic orbits in the attractor are saddles. The origin period-doubles as $\sigma$ increases at $\sigma = \sigma_0 = 1/\pi \sim 0.318$ (the HCB value).
As $\sigma$ increases from beyond $\sigma_0$ a new index-$2$ set appears in the attractor, and repelling periodic orbits are immediately dense in the attractor (Fig.~\ref{fig:Kos-saddles} left for $\sigma=0.35$), and the saddle periodic orbits are still dense in the attractor (Fig.~\ref{fig:Kos-saddles} right).

\ifincludeXX
\begin{figure}
\includegraphics[width=.238\textwidth,height=.238\textwidth]
{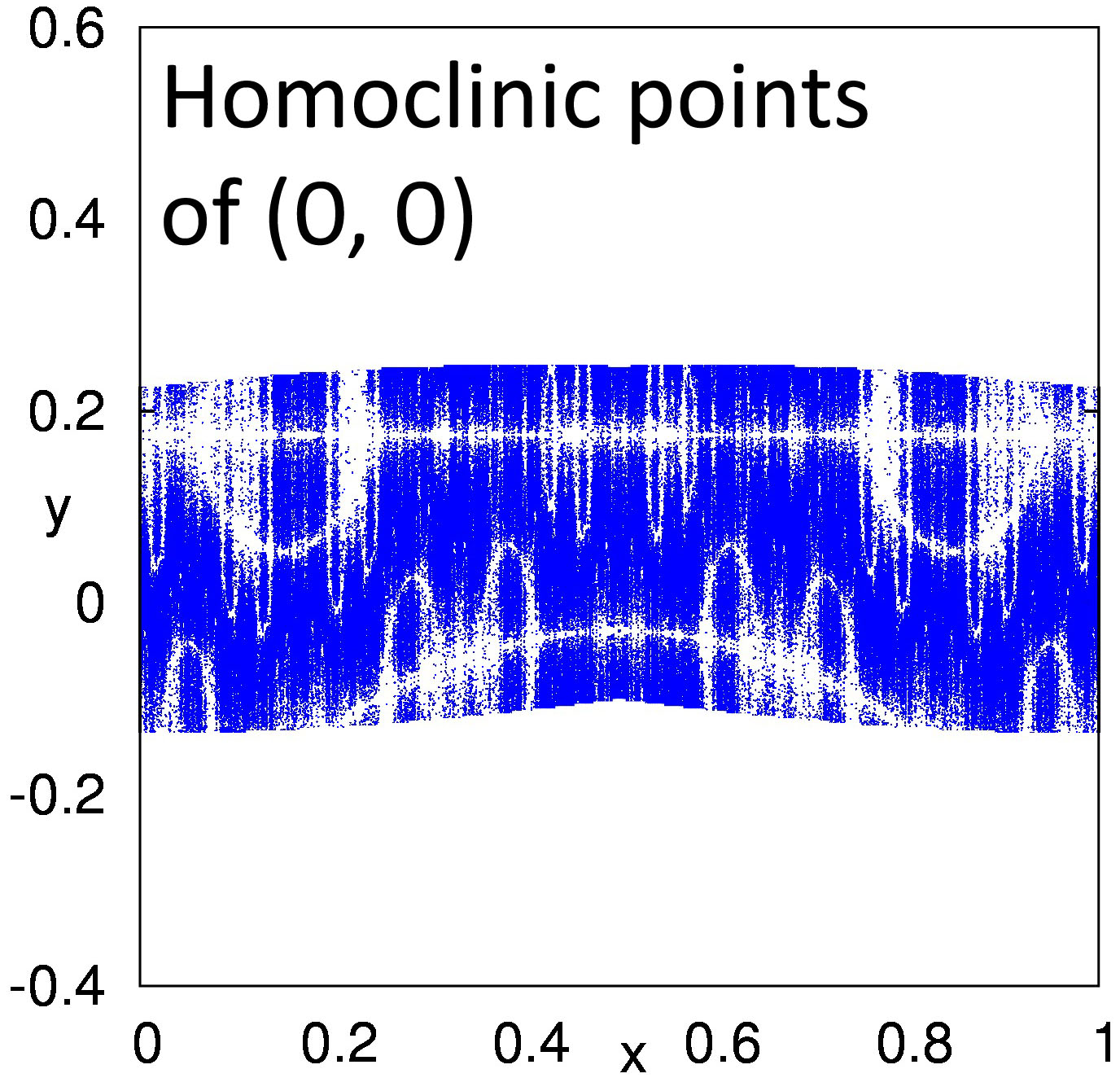}
\includegraphics[width=.238\textwidth,height=.238\textwidth]
{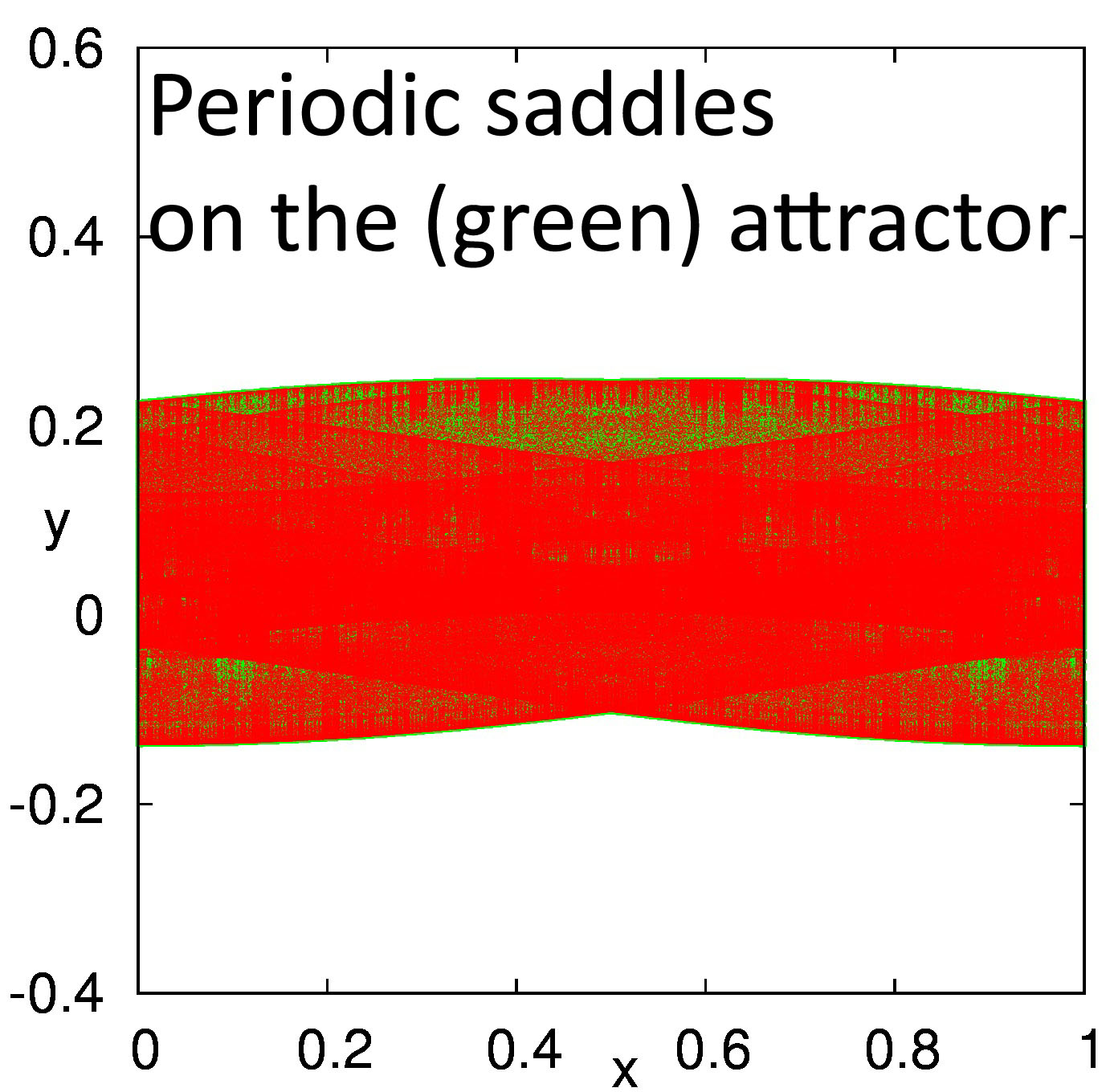}
\caption{ {\bf Homoclinic points and periodic saddles  for Kostelich map.}
It can be shown that both saddles and repellers are dense in the attractor, so that we have hetero-chaos. This figure shows what the sets of homoclinic points and periodic points look like for limited computations. 
{\bf Left panel.} 
Points in the attractor that map to the repelling origin within $14$ iterates. Since they are in the unstable manifold of $(0,0)$, they are homoclinic points.
 In fact, the homoclinic points can be shown to be densely distributed in the attractor implying that repelling periodic points are dense, since  Marotto \cite{marotto_2005} shows that arbitrarily close to each homoclinic point there are repelling periodic points.
 {\bf Right panel.} 
 The saddle periodic points (red) of 
period $13$ 
are plotted on top of the chaotic attractor (green).
They become denser as the period increases. 
}
\label{fig:Kos-saddles}
\end{figure}
\fi

{\bf Upper-triangular Jacobians.} 
Our hetero-chaos baker maps and the maps in this section have the following property. 
Each periodic orbit lying wholly in some $\R_k$ has UD-$k$. This is true because 
for each map $F$, the Jacobian matrix $DF(x,y)$ is lower triangular.
The Jacobian $DF^T$ of the time-$T$ map $F^T$ is also lower triangular since by the chain rule, $DF^T$ is the product of $T$ of these matrices $DF$. The number of expanding directions for a point is the number of diagonal elements of $DF$ that are $>1$.

\section{Lorenz-96 model.} 
So far in this paper we have considered maps rather than differential equations in order to keep the models as simple as possible, but our real goal is to understand higher dimensional hetero-chaotic differential equations.
Edward Lorenz proposed a variety of closely related chaotic differential equation models. See \cite{saiki_2017} for connections among them and for some generalizations. In particular 
Lorenz  \cite{lorenz_1996,lorenz_1998} proposed a dissipative $N$-dimensional ODE as a model of some oscillating scalar atmospheric quantity described by 
\begin{equation}\label{eqn:lorenz}
  \frac{dx_k}{dt}=x_{k-1}(x_{k+1}-x_{k-2})-x_k+F, \mbox{ for }k= 1,\ldots,N,
\end{equation}
where the system has cyclical symmetry so $x_{N+k} = x_k$ for all $k= 1,\ldots,N$, and 
where $F$ is a forcing parameter. We use the case $N=8$. 
For the Lorenz-96 model with $F=8$, 
the chaotic attractor has Lyapunov dimension 5.379. 
Numerically, we found many periodic orbits of UD-$1$, $2$, and $3$, and no periodic orbits with UD-$k$ ($k>3$).
Three of them with different UD values are shown in Fig.~ \ref{fig:orbit-in-chaos}.
\ifincludeXX
\begin{figure}
\includegraphics[width=0.22\textwidth]{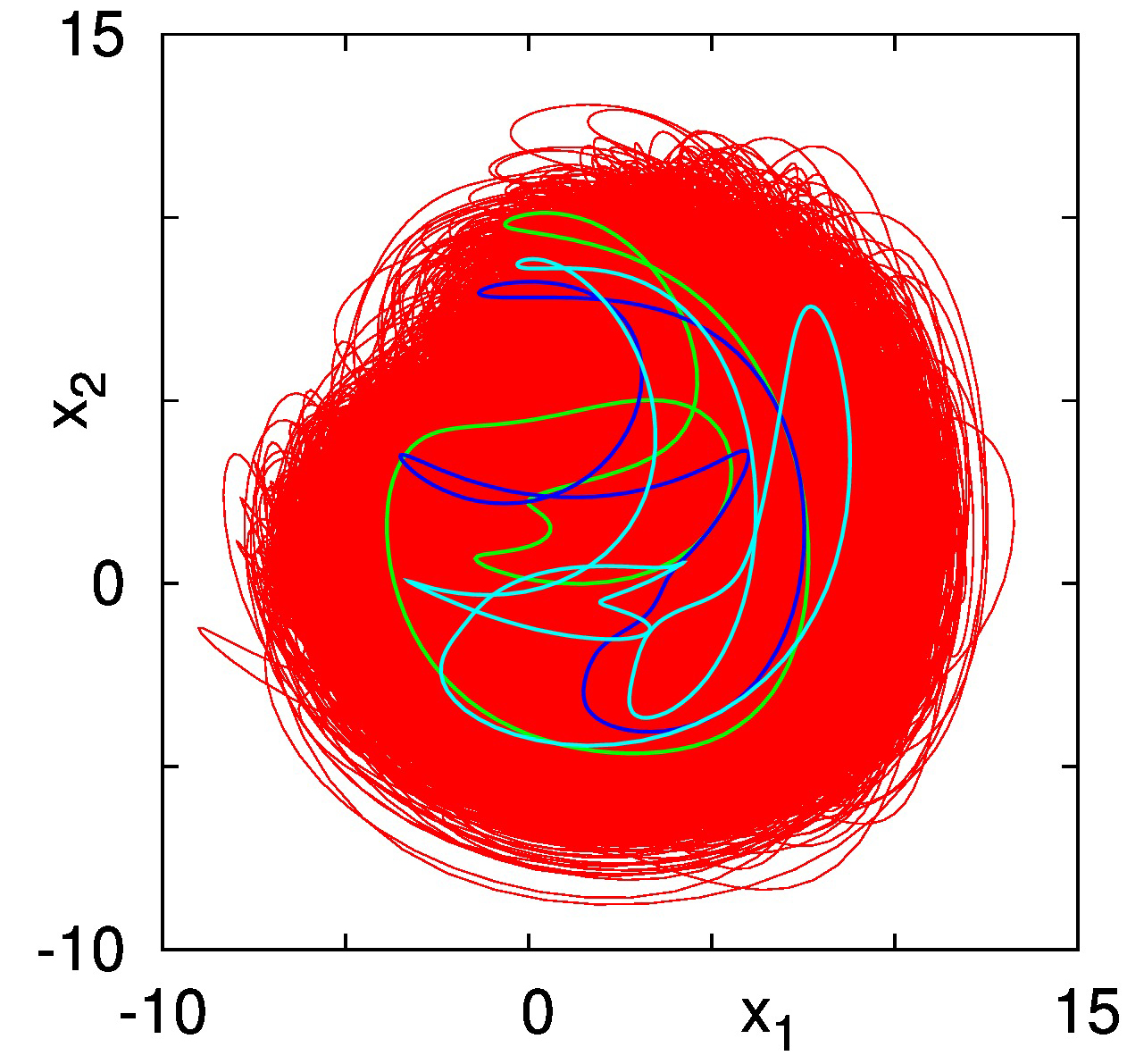}
	\caption{\textbf{A projection of the chaotic attractor and three periodic orbits with different UD values.}
	  This shows a projection into the $x_1-x_2$ plane of orbits $O_1, O_2,$ and $O_3$ with UD-1 (green), 2 (blue), and 3 (light blue), respectively. There are infinitely many possible projections of $\mathbb R^8$ into a plane and all those tested show all three periodic orbits lying within the projected attractor. See also Fig. \ref{approach-to-3-orbits}.}
\label{fig:orbit-in-chaos}
\end{figure}
\begin{figure}[h]
\includegraphics[width=.22\textwidth]{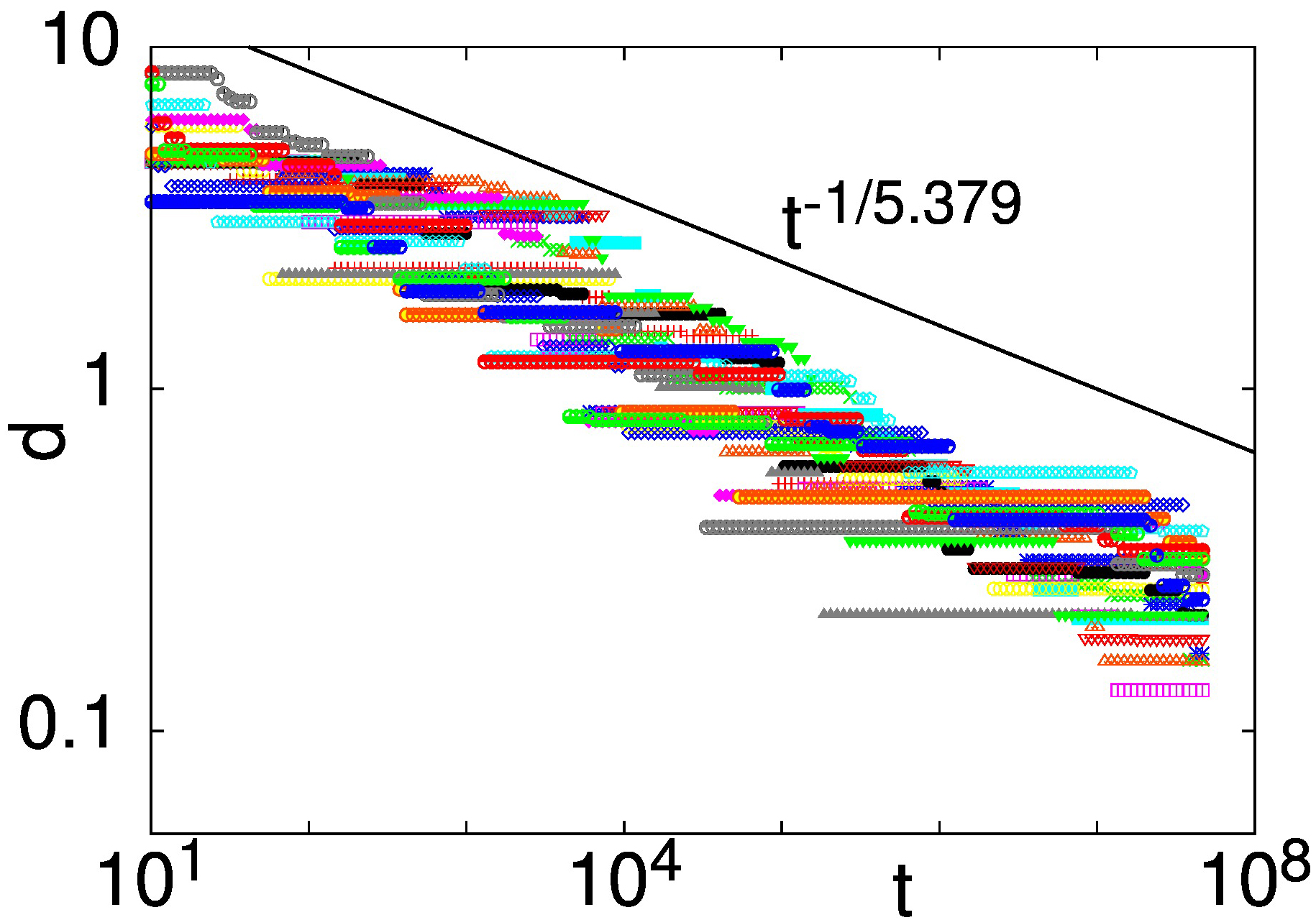}
\includegraphics[width=.22\textwidth]{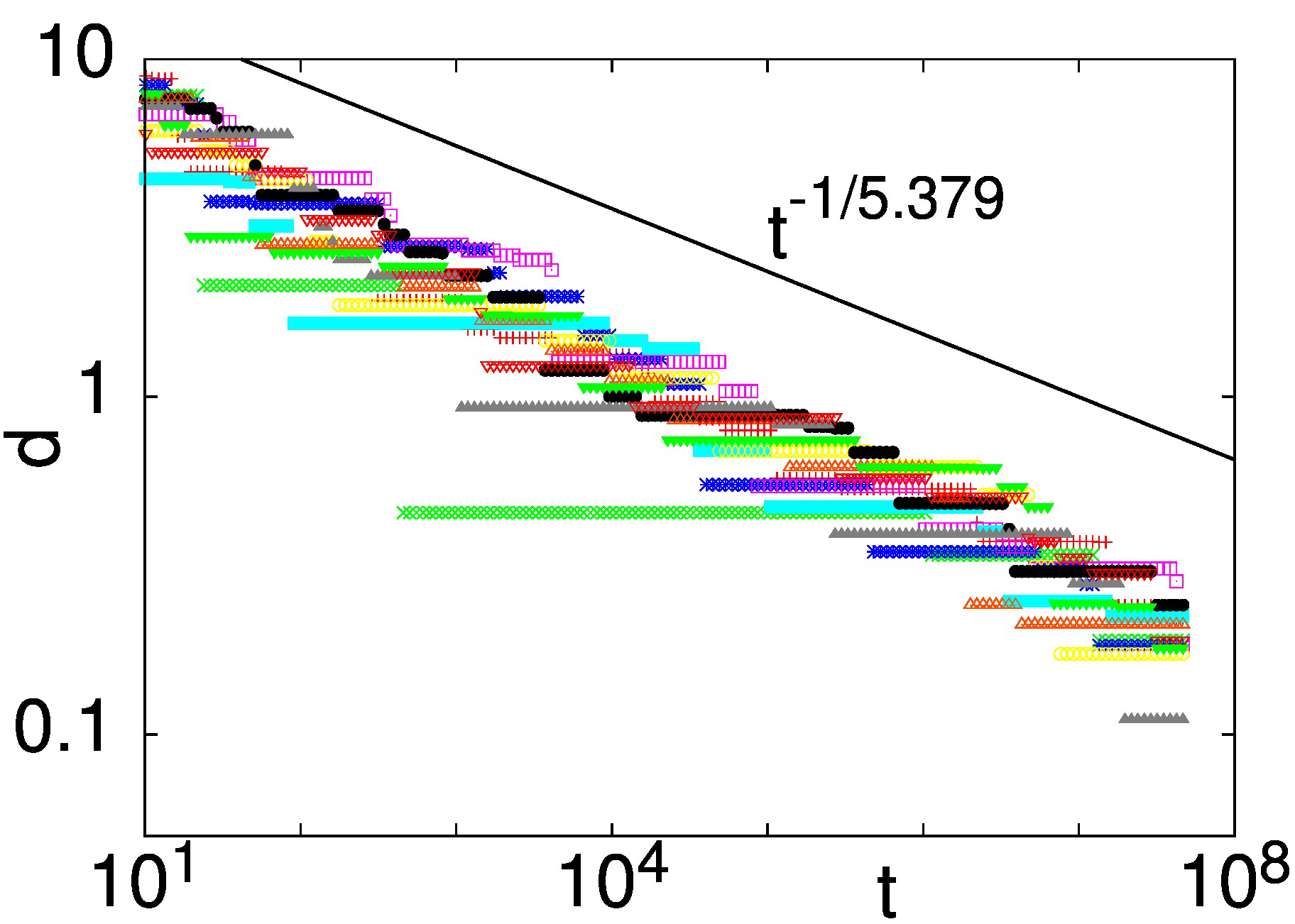}
\caption{
{\bf  Trajectories approach 3 periodic orbits.}
We investigated three periodic orbits  $O_1, O_2,$ and $O_3$ in 
Fig. \ref{fig:orbit-in-chaos}, $O_1$ in the left panel and $O_2$ in the right. The graph for $O_3$ is omitted since it is quite similar to the right panel. 
We chose 21 initial conditions very close to each other.
In each panel, for each of 21 trajectories, the closest approach by time $t$ to the respective periodic orbit is reported. 
The 8 Lyapunov exponents for the global chaotic attractor are as follows: 
$1.594$, $0.390$,  $0.0$, $-0.453$, $-0.960$, $-1.508$, $-2.450$, $-4.613.$
A standard estimate of a dimension of a chaotic attractor is its ``Lyapunov'' dimension \cite{alligood_1996}.
The Lyapunov dimension of the attractor is $5.379$ and according to a long-standing conjecture, \textcite{farmer_1983} the closest approach of a typical trajectory to a typical point of the attractor is expected to be proportional to $t^{-1/5.379}$ for time $t\in[0,t]$.
The straight line indicates that rate of closest approach to that orbit (though periodic orbit points are not generally typical). The straight line has been shifted vertically slightly.  For all three, the actual convergence appears slightly faster than expected. It suggests that given sufficient time the typical trajectory would come arbitrarily close to all three periodic orbits.
The Lyapunov dimension for these periodic orbits are 5.268 for $O_1$,
5.107 for $O_2$, and 5.514 for $O_3$.
}
\label{approach-to-3-orbits} 
\end{figure}
\fi
The distances between three periodic orbits and chaotic orbits were calculated and shown in Fig. \ref{approach-to-3-orbits}.
This implies that the periodic orbits are in the attractor and the attractor has UDV, and so also is hetero-chaotic according to our conjecture.

\section{\bf Discussion} 
Hetero-chaos is important for all models with high-dimensional attractors including weather prediction and climate modeling. 
It is perhaps the unifying concept linking different phenomena observed in numerous numerical simulations of chaotic dynamical systems and physical experiments, such as unstable dimension variability (UDV), on-off intermittency, riddled basins, blowout and bubbling bifurcations. It is also a major cause of shadowing to fail, i.e., for simulated solutions to be non-physical. We have made three conjectures as the beginning of a general theory of hetero-chaos.

Hetero-chaotic systems are particularly difficult to visualize, so we have introduced some low-dimensional examples as paradigms, 
including one that is perhaps the simplest possible example of hetero-chaos (based on the well-known baker map). See  Figs.~\ref{fig:2D_HC_baker} and ~\ref{fig:3D_HC_baker}.

We investigate how hetero-chaos arises as a parameter is varied.
It can either occur at a crisis, that is a sudden jump in the size of the chaotic attractor, or it can occur when the attractor is changing continuously. In such cases we find that the transition to hetero-chaos occurs at a periodic orbit bifurcation, and we believe this is the typical case when the attractor varies continuously.
Because shadowing fails for hetero-chaotic systems, detecting the transition from homogeneous chaos to hetero-chaos can be critical for prediction efforts. 

While the UDV condition requires only two orbits of different UD values,
we have focused on the existence of not just these two orbits but much larger index sets which exist in hetero-chaotic attractors and make hetero-chaos persistent.

Because of the increasing importance of models with high dimensional chaotic attractors, we have tried to create terminology that is easy to use.

\FloatBarrier 

\begin{acknowledgments}
YS was supported by the JSPS KAKENHI Grant No.17K05360 and JST PRESTO JPMJPR16E5.
MAFS was supported by the Spanish State Research Agency (AEI) and the European Regional Development Fund (FEDER)
No.FIS2016-76883-P and jointly by the Fulbright Program and the Spanish Ministry of Education
No.FMECD-ST-2016.
\end{acknowledgments}

\end{document}